# Classification of Diabetes Mellitus using Modified Particle Swarm Optimization and Least Squares Support Vector Machine

Omar S. Soliman[1], Eman AboElhamd[2]

*1,2 Faculty of Computers and Information, Cairo University,
5 Ahmed Zewail Street, Orman, Giza, Egypt*

**ABSTRACT:** *Diabetes Mellitus is a major health problem all over the world. Many classification algorithms have been applied for its diagnoses and treatment. In this paper, a hybrid algorithm of Modified-Particle Swarm Optimization and Least Squares-Support Vector Machine is proposed for the classification of type II DM patients. LS-SVM algorithm is used for classification by finding optimal hyper-plane which separates various classes. Since LS-SVM is so sensitive to the changes of its parameter values, Modified-PSO algorithm is used as an optimization technique for LS-SVM parameters. This will Guarantee the robustness of the hybrid algorithm by searching for the optimal values for LS-SVM parameters. The pro-posed Algorithm is implemented and evaluated using Pima Indians Diabetes Data set from UCI repository of machine learning databases. It is also compared with different classifier algorithms which were applied on the same database. The experimental results showed the superiority of the proposed algorithm which could achieve an average classification accuracy of 97.833%.*

## 1. INTRODUCTION

Diabetes Mellitus, or simply Diabetes, is a group of metabolic diseases in which a person has high blood sugar, either because the pancreas does not produce enough insulin, or because cells don't respond to the insulin that is produced. This high blood sugar produces the classical symptoms of polyuria (frequent urination), polydipsia (increased thirst) and polyphagia (increased hunger). There are 3 major types of DM, "Type I DM", which results from the body's failure to produce insulin, and currently requires the person to inject insulin or wear an insulin pump. This form was previously referred to as "insulin-dependent diabetes mellitus" (IDDM). The second type of DM is called "Type II DM" which results from insulin resistance, a condition in which cells fail to use insulin properly, sometimes combined with an absolute insulin deficiency. This type also named as "Non insulin-dependent diabetes mellitus" (NIDDM) or "adult-onset diabetes". Finally, "gestational diabetes" occurs when pregnant women without a previous diagnosis of diabetes develop a high blood glucose level, It may precede development of type I DM. Other forms of DM include congenital diabetes, which is due to genetic defects of insulin secretion, cystic fibrosis-related diabetes, steroid diabetes induced by high doses of glucocorticoids, and several forms of monogenic diabetes [1], [2], [4].

All types of DM have something in common. Normally, your body breaks down the sugars and carbohydrates you eat into a special sugar called glucose. Glucose fuels the cells in your body. But the cells need insulin, a hormone, in your bloodstream in order to take in the glucose and use it for energy. All types of DM have been treatable since insulin became available in 1921. Both type I & II are chronic conditions that cannot be cured. pancreas transplants have been tried with limited success in type I DM, gastric bypass surgery has been successful in many with morbid obesity and type II DM. Gestational DM usually resolves after delivery [10]. Untreated DM can cause many complications. Acute complications include diabetic ketoacidosis and non ketotic hyperosmolarcoma. Series long term complications include cardiocascular disease, chronic renal failure, and diabetic retinopathy. Adequate treatment of the disease is very important, as well as blood pressure control and lifestyle factors such as stopping smoking and maintaining a healthy body weight. Since the cells can't take in the glucose, it builds up in your blood. High levels of blood glucose can damage the tiny blood vessels in your kidneys, heart, eyes or nervous system. that's why diabetes can eventually cause heart disease, stroke, kidney disease, blindness and nerve damage to nerves in the feet (especially if left untreated) [8], [26]. The aim of this paper is to develop a classification algorithm for DM diagnosis and treatment using a hybrid algorithm consists of Modified-PSO algorithm and LS-SVM classifier.

Least Squares-Support Vector Machine (LS-SVM) classifier is one particular sample of Support Vector Machine (SVM) [24]. LS-SVM is used for finding an optimal hyper plane, which separates various classes. It obtains this optimal hyper-plane by using maximum Euclidean distance to the nearest point. It is a parametric algorithm that is popular with its sensitivity to the changes in the values of its parameters. Particle Swarm Optimization (PSO) is a heuristic algorithm inspired from the nature social behavior of birds. The main strength of PSO is its fast convergence, compared with other global optimization algorithms [6]. The rest of this paper is organized as follows. Section 2 describes the problem background and related work. The proposed hybrid algorithm is introduced in section 3 while experimental results are presented in section 4. The last section is devoted to the conclusion and further research.

## 2. BACKGROUND AND RELATED WORK

Diabetes Mellitus (DM) is a major health problem in both industrial and developing countries and its incidence is rising [1], [2]. Many classification algorithms have been applied on this area trying to classify the patients or predict their future state. This section will introduce some of these works.

A brief review and discussion of the philosophy, capabilities, and limitations of Artificial Neural Network (ANN) in medical diagnosis through selected examples including DM was introduced in [3]. A hybrid binary classification model using the basic concepts of soft computing and ANN was proposed in [5]. A novel machine learning algorithm termed "mixture of expert" was used for the determination of a patient's diabetic state [7]. A model using ANN with RBF kernel and one hidden layer was proposed in [27]. The artificial metaplasticity on multilayer perceptron (AMMLP) was used as prediction model for diabetes. The best result obtained from AMMLP algorithm was 89.93% [12]. In [30] evaluation for the efficacy and safety of canagliflozin in subjects with T2DM and stage 3 chronic kidney disease was made using different classification techniques. ANN and multivariate logistic regression (MLR) models was proposed in [28]. A multilayer perception NN and a conditional logistic regression were used to predict albuminuria in type II DM [14]. Two statistical models were used to predict albuminuria in type II DM in [15]. A survey of more than one supervised and unsupervised algorithms was introduced in [20]. SVM technique was proposed for classification of DM patients. The results showed a sensitivity of 99.45% for the classifier and specificity of 100% [18]. A classification Algorithm based on Fuzzy systems, Evolutionary Algorithms (ACO) and ANN techniques was proposed [13].

In [7], Six different neural networks (Probabilistic Neural Network (PNN), Learning Vector Quantization (LVQ), Feed-Forward Networks (FFN), Cascade-Forward Networks (CFN),Distributed Time Delay Net-works (DTDN), Time Delay Networks (TDN)), artificial immune system and Gini algorithm from decision tree were used for DM patients classification. A comparative study of different classification techniques had been done in [19]. A study which worked on 1506 participants was held. The main outcome measures were age specific mortality rates due to cardiovascular disease and all causes [16]. In [25] A robust version of Support Vector Machine (SVM) based on Value-at-Risk (VaR) measure referred to as VaR-SVM was proposed. A hybrid model that integrates Genetic Algorithm and Back Propagation network (BPN) was proposed in [9]. A hybrid binary classification model was proposed for diabetes type II classification, based on the basic concepts of soft computing and artificial intelligence techniques [11] .

*2.1 Modified Particle Swarm Optimization*

Particle Swarm Optimization (PSO) is an algorithm Inspired from the nature social behavior and dynamic movements and communications of insects, birds and fish [6], [23]. The main strength of PSO is its fast convergence, comparing with many global optimization algorithms like Genetic Algorithms (GA), Simulated Annealing (SA) and other global optimization algorithms. The key concept is dealing with changes in velocity. In general, the main idea of PSO is as follows. For the $i^{th}$ particle in d dimension, it could update its velocity and position using, (1) and,(2). Where $r_1$ and $r_2$ are two random numbers in the range[0, 1], $V_{id}$ is the momentum, $\omega_{id}$ is the interia weight, $C_1$ is the cognitive learning parameter and $C_2$ is the social collaboration parameter. $X_{id} = (x_{i1}, x_{i2}, \ldots, x_{id})$ is the position of the $i^{th}$ particle, $P_i = (p_{i1}, p, \ldots, p_{id})$ represents the best previous position (i.e. the position with the highest fitness value).

$$V_{id} = \omega_{id}V_{id} + C_1r_1(p_{id} - X_{id}) + C_2r_2(p_{gd} - X_{id}) \quad (1)$$

$$X_{id} = X_{id} + V_{id} \quad (2)$$

Inertia Weight plays an important role in the process of providing balance between exploration and exploitation. It determines the contribution rate of a particles previous velocity to its velocity at the current time step. In [5] different types of inertia weights were mentioned like Constant, Random, Adaptive inertia weight and many other types. In [29] a modified version of PSO was proposed, The main idea of this modified version is as in the following equations. For the $i^{th}$ particle in $d$ dimention, it could update its velocity and position using,(3) and,(4)

$$V_{id} = \lambda[\omega_{id}V_{id} + C_1r_1(p_{id} - X_{id}) + C_2r_2(p_{gd} - X_{id})] \quad (3)$$

$$X_{id} = X_{id} + (\omega V_{id}) \quad (4)$$

where $\lambda$ is a convergence factor, which can be calculated using, (5)

$$\lambda = \frac{2}{|2 - C - \sqrt{C^2 - 4C}|} \quad (5)$$

**Where $C = C_1 + C_2$**

In the proposed Algorithm $\omega_{id}$ could be calculated using,(6) where t is the iterator over all iterations and $T_{max}$ is the maximum number of iterations. With the increasing of $t$, parameter $\omega$ will be decreased linearly from 0.9 to 0.4 [5].

$$\omega_{id} = 0.9 - \frac{t}{T_{max}} * 0.5 \quad (6)$$

The Modified-PSO algorithm steps is illustrated in Algorithm-1 with random inertia weight [29].

_______________________________________________

*Algorithm-1: Modified-PSO*

**Step 1:** Initialize population of particles $X(t)$ which consists of random positions $x_1, x_2, \ldots, x_n$ and velocities $V(t)$ are made up of the particle's initial velocity $v_1, v_2, \ldots, v_n$ on n dimensions.

**Step 2:** Evaluate the fitness for each particle.

**Step 3:** For each particle, find the maximum fitness and compare it to the best found so far (pbest), if $f(x_i) < f(pbest_k)$, then $f(pbest_k) = x_i$

**Step 4:** Set $P_i$ equals to the location of the maximum fitness value $X_i$

**Step 5:** Compare fitness evaluation with the population's overall previous best. If current value is better than $gbest$, then reset $gbest$ to the current particle's array index and value.

**Step 6:** Calculate the convergence factor $\lambda$ using,(5)

**Step 7:** Calculate the Inertia weight $\omega_{id}$ using,(6)

**Step 8:** Update the position of the particle according to, (3) and, (4) and the new population $X(t+1)$ will be generated.

**Step 9:** Adjust the acceleration of the particles using, (7)

$$v_i = \begin{cases} V_{max} & if\ v_i > V_{max} \\ -V_{max} & if\ v_i < -V_{max} \end{cases} \quad (7)$$

**Step 10:** Loop to step (2) until stopping criterion is satisfied (Reach a maximum number of iterations $T_{max}$)

__________________________________________________

*2.2 Least Squares Support Vector Machine*

Least Squares-Support Vector Machine (LS-SVM) classifier is one particular sample of Support Vector Machine (SVM) [31], [24]. One could finds the solution in LS-SVM by solving a set of linear equations instead of a convex quadratic programming problem for classical SVMs, The main target of LS-SVM is finding an optimal hyper plane, which separates various classes. It obtains this optimal hyper-plane by using maximum Euclidean distance to the nearest point. The LS-SVM classifier maps the input vectors into a high dimensional feature space for non-separable data. Then, the LS-SVM classifier finds an optimal separating hyper-plane in this higher dimensional space [22].

Given a training dataset of $N$ points $\{x_k, y_k\}_{k=1}^N$ with input data $x_k \epsilon R^n$ and output $y_k \epsilon R$, we consider the following optimization problem in primal weight space:

$$\min J(w,b)_{w,b,e} = \frac{1}{2}w^T w + \frac{1}{2}\gamma \sum_{k=1}^N e_k^2 \quad (8)$$

such that

$$y_k - (w^T \varphi x_k + b) = e_k, k = 1,2,\ldots N \quad (9)$$

Where $\gamma$ is a regularization factor, $e_k$ the difference between the desired output $y_k$ and the actual output, and $\varphi(.)$ is a nonlinear function mapping the data points into a high dimensional Hilbert space; in addition, the dot product in the high-dimensional space is equivalent to a positive definite kernel function $K(x_i, x_j) = \varphi(x_i)^T \varphi(x_j)$. In primal weight space, a linear classifier in the new space takes the following form Where $w$ is the weight vector and $b \epsilon R$ which called as the bias term.

$$y(x) = sign(w.\varphi(x) + b) \quad (10)$$

The dual space of this primal space was found by solving the Lagrangian function in, (11)

$$L(w,e,\propto) = J(w,e) - \sum_{k=1}^N \propto_k (w^T \varphi(x_k) + e_k - y_k) \quad (11)$$

Where $\propto_k$ are Lagrangian multipliers and are called Support Vectors. The optimal solution for objective function in, (11) must satisfy the following Karush-Kuhn Tucker (KKT) conditions [22].

$$\frac{\delta L}{\delta w} = 0 \rightarrow w = \sum_{k=1}^N \alpha_k y_k \varphi(x_k) \quad (12)$$

$$\frac{\delta L}{\delta w} = 0 \rightarrow \propto_k = \gamma e_k, \ k = 1,\ldots,N$$

$$\frac{\delta L}{\delta w} = 0 \rightarrow w^T \varphi(x_k) + e_k - y_k = 0, k = 1,\ldots N$$

The linear system in, (13) will results after elimination of w and e which generates the Support Vector $\propto'_k$

$$\left(K + \frac{I}{\sigma}\right)\alpha = y \quad (13)$$

Where $y = [y_1, y_2, \ldots, y_N]^T$, $\propto = [\propto, \propto_2, \ldots, \propto_N]^T$ and $K \epsilon R^{NxN}$ is the kernel matrix. The resulting LS-SVM model for function estimation is as in, (14) where $K(.,.)$ is the kernel function.

$$y(x) = \sum_{k=1}^N \propto_k K(x, x_k) \quad (14)$$

LS-SVM (Algorithm-2) was implemented using Ra-dial Basis Function (RBF), (15) [22].

$$K(x, x_k) = \exp(-\frac{|x-x_k|^2}{\sigma^2}) \quad (15)$$

__________________________________________________

*Algorithm-2: LS-SVM:*

**Step 1:** Load the training data set of n data points, $\{x_k, y_k\}_{k=1}^N$ where $x_i$ is the $i^{th}$ input vector and $y_i \epsilon R$ is the corresponding $i^{th}$ target with values $\{-1, +1\}$.

**Step 2:** Generate random weights for each input data point.

**Step 3:** Determine the value of the bias term b and initialize the error e for each point randomly.

**Step 4:** Initialize $\gamma$ and $\sigma$ using random values.

**Step 5:** Search for values of e, w and b that minimize the objective function, (8) and, (9).

**Step 6:** Construct the Lagrangian function in, (11) with the solution that must satisfy the KKT conditions in the set of, (12).

**Step 7:** Calculate number of support vectors($\propto$) using, (13).

**Step 8:** Training data for LS-SVM model could be classified using, (14) with RBF kernel function, (15).

**Step 9:** Classify any new point by, (10) using RBF kernel function, (15) .

**Step 10:** Loop until stopping criteria is met, usually until reach the maximum number of iterations.

__________________________________________________

The proposed algorithm worked on 768 record from Pima Indians Diabetes Data set which contains 8 features as shown in TABLE (I). Among those 768 cases, there are 500 healthy case and 268 suffered from DM.

TABLE I. PIMA INDIANSDIABETESDATA SET

| Feature Number | Feature Name |
|---|---|
| 1 | Number of times pregnant. |
| 2 | Plasma glucose concentration a 2 h in an oral glucose tolerance test. |
| 3 | Diastolic blood pressure. |
| 4 | Triceps skin fold thickness. |
| 5 | 2-h serum insulin. |
| 6 | Body mass index. |
| 7 | Diabetes pedigree function |
| 8 | Age |
| Class Label | 0 if patient and 1 if healthy |

### 3. PROPOSED ALGORITHM

In light of the previous literature review and background, A hybrid classification algorithm which integrates Modified- PSO algorithm as a parameters optimization technique and LS-SVM for classification was proposed.

The proposed algorithm for DM diagnosis and treatment is composed of two main phases, Parameters Optimization and Classification. Modified-PSO algorithm was used as a parameters optimization technique aiming to improve the sitting of the parameter values of LS-SVM. Hence, overcoming it's sensitivity to the parameter values changes. Classification phase using LS-SVM technique consists of two main phases, Training phase followed by a Testing phase. A block diagram of the algorithm is Depicted in Fig(1).

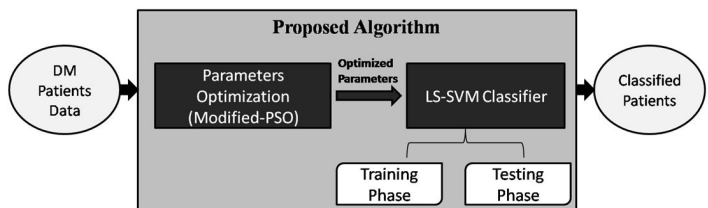

Fig. 1. Block diagram of the proposed algorithm

The aim of parameters optimization phase using Modified-PSO (Section 2.1) is to find the optimal values for the parameters of the LS-SVM classifier (The regularization factor ($\sigma$) and Gaussian Kernel function ($\gamma$). The second phase utilizes LS-SVM to classify the DM patients into one of two classes (Live/Die) using the optimized parameters. Algorithm-3 illustrates the proposed algorithm in details.

______________________________________________

***Algorithm-3: Proposed Algorithm:***

**Step 1:** Load the data set of n data points, $\{x_k, y_k\}^N_{k=1}$ where $x_i$ is the $i^{th}$ input vector and $y_i \epsilon R$ is the corresponding $i^{th}$ target with values $\{-1, +1\}$.

**Step 2:** Generate random weights for each input data point.

**Step 3:** Initialize the bias term b and the error e for each point randomly.

**Step 4:** Find the optimal values for $\gamma$ and $\sigma$ using algorithm 1.

**Step 5:** Find optimal values of $(e, w\ and\ b)$ for the objective function in, (8) and, (9).

**Step 6:** Calculate number of support vectors ($\propto$)using, (13)

**Step 7:** Classify any new point by, (10) using RBF kernel function, (15)

**Step 8:** Loop until stopping criteria is met, usually until reach the maximum number of iterations.

______________________________________________

### 4. EXPERIMENTAL RESULTS

As mentioned before, the proposed algorithm worked on Pima Indians Diabetes Data Set. Some of the records in this data set contains zeros in some features. Cells with zero entry were kept unchanged (neither deleted nor filled) [13]. The input for the Modified-PSO algorithm is total of 768 records. About 768 random individuals are generated in the search space for 100 Iterations. The output from the Modified-PSO is the optimal values for $\gamma$ and $\sigma$, which are 100 and 0.5 respectively. LS-SVM classifier is run with the optimized parameters and RBF kernel function, (15), seeking to find the optimal hyperplan that separates the search space into 2 classes (Live, Die) by finding the optimal values for (w, e and b) in the objective function, (8) and, (9)

In order to evaluate the performance of the proposed Algorithm, the classification accuracy was calculated using, (16). Where TP and TN stand for True Positive and True Negative which are the proportion of positive and negative cases that were correctly identified respectively. Positive cases are the records with Live label and negative ones are with Die label. FP and FN stand for False Positive and False Negative which are the proportion of negative cases that were incorrectly classified as positive and the proportion of positive cases that were incorrectly classified as negative respectively [13].

$$Accuracy = \frac{TP+TN}{TP+TN+FP+FN} \qquad (16)$$

The training phase was implemented using 10-fold Cross Validation (CV) method which breaks the data set into 10 sets of size n/10, train on 9 data sets and test on 1, then repeat this process 10 times and take a mean accuracy [17]. The mean classification accuracy for LS-SVM is 97.833%, which obtained from the RBF kernel function, (15). TABLE (II) shows the accuracy for every fold while applying 10-fold CV, It's obvious that the average accuracy over all folds is 97.833%.

TABLE II. ACCURACY OF 10-FOLD CV

| Fold | Accuracy |
|---|---|
| 1 | 93.993% |
| 2 | 95.973% |
| 3 | 96.889% |
| 4 | 99.9769% |
| 5 | 97.991% |
| 6 | 98.698% |
| 7 | 96.999% |
| 8 | 99.988% |
| 9 | 99.99% |
| 10 | 97.83% |

The proposed Algorithm is compared with different recently classifier Algorithms which were applied on the same database, as shown in TABLE (III). These techniques are multilayer perceptron (AMMLP) [12], Fuzzy, Decision Tree, ACS and ANN [13], 6 different ANN types and AIS technique [7], Also, GA and ANN [9], MLP and other techniques [11] and Different Evolutionary algorithms with different tools [19].

TABLE III. AVERAGECLASSIFICATIONACCURACY OF THE PROPOSED ALGORITHM AND OTHER CLASSIFCATION ALGORITHMS WITH THE NUMBER OF USED RECORDS

| CI Technique | Accuracy | Diagnosed Patients |
|---|---|---|
| ANN and AIS [7] | 76% | 768 Patients |
| MLP/BN/J48graft/JRip and FLR [19] | 81.33% | 768 Patients |
| MLP, SVM, KNN,QDA and LDA [11] | 82.4% | 768 Patients |
| GA and ANN [9] | 84.713% | 392 Patients |
| AMMLP [12] | 89.93% | 768 Patients |
| Fuzzy, DT, ACS and ANN [13] | 95.852% | 247 Patients |
| **The Proposed Algorithm** | **97.833%** | **768 Patients** |

Fig(2) demonstrates the results of the average classification accuracy of the proposed algorithm against the accuracy of other classification algorithms applied on the same data set. Results show the effectiveness of the proposed Algorithm, which has the maximum average classification accuracy of 97.833% over other algorithms. TABLE(III) also articulates the average classification accuracy obtained from all algorithms in addition to the number of records used from Pima Indians Diabetes Data set in each algorithm.

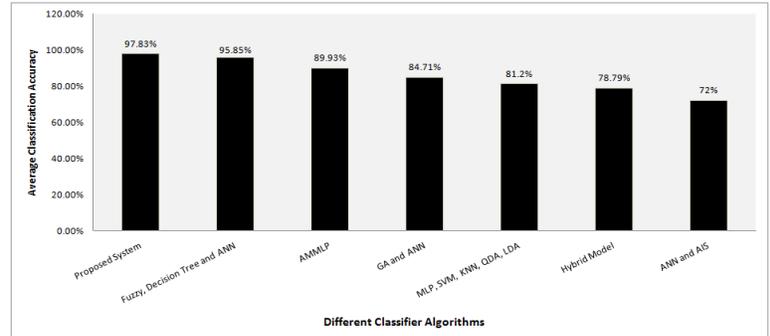

Fig. 2. Average Classification Accuracy of DM using the proposed algorithm and other classification algorithms

### 5. CONCLUSION AND FUTURE WORK

This paper introduced a hybrid classification Algorithm for DM patients. The proposed Algorithm integrates modified version of PSO and LS-SVM algorithm. The proposed algorithm was composed of two main phases which are Parameters Optimization and Classification. Classification had two main phases, Training phase followed by a phase of Testing the algorithm. The input parameters for LS-SVM were optimized using modified version of PSO algorithm. The LS-SVM algorithm was used to classify DM patients into one of two classes (Live/Die). Modified-PSO could guarantee the robustness of the hybrid algorithm by searching for the optimal values for LS-SVM parameters. Optimizing the parameters could minimize the classification time by avoiding making trial and error while targeting the optimal values for LS-SVM parameters. The proposed algorithm was implemented on Pima Indians Diabetes Data set from UCI repository of machine learning databases. The average classification accuracy of LS-SVM method with RBF kernel was 97.833% which is the best while compared with other algorithms which worked on the same data set. As a future work, Ant Colony System (ACS) could be used as an optimization technique. Also, other kernel functions could be applied in the classification phase.